\documentstyle[twoside,fleqn,espcrc2]{article}

\newcommand{\AmS}{{\protect\the\textfont2
  A\kern-.1667em\lower.5ex\hbox{M}\kern-.125emS}}

\title{Double parton scatterings in high energy hadronic collisions}

\author{G. Calucci and D. Treleani\address{Dipartimento di Fisica Teorica
dell'Universit\`a and INFN, Sezione di Trieste \\ 
        Trieste, I 34014 Italy}%
        \thanks{This work was partially supported by the Italian Ministry of
        University and of
Scientific and Technological Research by means of the Fondi per la Ricerca
scientifica - Universit\`a di Trieste.}
       }
       
\begin{document}

\begin{abstract}
CDF has recently measured a large number of double parton scatterings.
The observed value of $\sigma_{eff}$, the non perturbative 
parameter which characterizes the process, is considerably smaller as 
compared with the naive
expectation. The 
small value of $\sigma_{eff}$ is likely to be an indication of the importance
of the two-body parton correlations in the
many-body parton distributions of the proton.
\end{abstract}

\maketitle

\section{INTRODUCTION}

The inclusive cross section for a double parton scattering,
namely of an event where, in the same inelastic interaction, 
two different pairs of partons scatter independently with large
momentum transfer, 
is written as\cite{Double}:

\begin{eqnarray}
\sigma_D=\int_{p_t^c}&D_2(x_A,x_A';{\bf b}) \hat{\sigma}(x_A,x_B)\nonumber\\
         &\hat{\sigma}(x_A',x_B') D_2(x_B,x_B';{\bf b})\\
         &d{\bf b}dx_Adx_Bdx_A'dx_B'\nonumber
\end{eqnarray}
 
\par\noindent
$ \hat{\sigma}(x_A,x_B)$ is the 
parton-parton cross section integrated with the cut off $p_t^c$,
which is the lower 
threshold to observe final state partons as minijets, 
$x$ is the momentum fraction, $A$ and $B$ are labels to identify the two 
interacting 
hadrons. 
$\sigma_D$ is a function of the product $\hat{\sigma}(x_A,x_B)
\hat{\sigma}(x_A',x_B')$. Actually the two different partonic interactions are 
localized in two regions in transverse space with a size of order $(1/p_t^c)^2$
and at a relative distance of the order of the hadronic radius $r$,
in such a way that the two
partonic interactions add incoherently in the double scattering
cross section. The non perturbative input in Eq.(1)
is the two-body parton distribution $D_2(x,x';{\bf b})$, which
depends on the fractional momenta of the two partons taking part to the 
interaction and on their
relative transverse distance ${\bf b}$. The transverse distance
${\bf b}$ has to be the same for
the two partons of hadron $A$ and the two partons of hadron $B$, in order to 
have
the alinement which is needed for a
double collision to occur. $D_2$ is a dimensional quantity and therefore the 
process introduces a
non perturbative scale factor which is related to the hadronic transverse size. 
\par\noindent
The simplest possibility to consider is the one where the dependence of 
$D_2$ on
the different variables is factorized:

\begin{equation}
D_2(x,x';{\bf b})=f_{eff}(x)f_{eff}(x')F({\bf b})
\end{equation}

\noindent
$f_{eff}$ is the effective parton distribution, namely the
gluon plus $4/9$ of the quark and anti-quark distributions
and $F({\bf b})$ is normalized to one. Multiparton
distribution are then uncorrelated and $D_2$ 
does not contain further informations
with respect to the one-body parton distribution (actually $f_{eff}$) 
apart form the dependence on ${\bf b}$, whose origin is the dimensionality of 
$D_2$ and which
gives rise to the scale factor $\sigma_{eff}$.
In fact in this case one may write

\begin{equation}
\sigma_D={\sigma_S^2\over\sigma_{eff}}
\end{equation}

\noindent
with

\begin{equation}
{1\over\sigma_{eff}}=\int F^2({\bf b})d^2b
\end{equation}

\noindent
and

\begin{equation}
\sigma_S=\int_{p_t^c}f_{eff}(x_A)f_{eff}(x_B)
  \hat{\sigma}(x_A,x_B)dx_Adx_B,
\end{equation}

\noindent
the single scattering expression of the perturbative QCD
parton model.
\par
Eq.(2) is the basic hypothesis
underlying the signature of a double parton collision which one has been
looking for experimentally\cite{CDF,Dexp}. 
The expected characteristic feature of a double collision 
is in fact that it should produce a final state analogous 
to the one obtained by super-posing two single scattering processes.
By looking at the dependence of $\sigma_{eff}$ on $x$ 
CDF has been able to verify the correctness of  
the factorization hypothesis in Eq.(2). The range of values of $x$ available
is limited to $x\le.2$, for the interaction producing a pair of minijets, and 
to $x\le.4$ for the 
interaction giving rise to a minijet and a photon.  
In the limited range of values of $x$ available,
the factorization hypothesis has shown to be consistent with the experimental 
evidence. 
\par	Since the uncorrelation hypothesis 
does not contradict the experiment,
one can work out the case where all multiparton distributions are 
uncorrelated and one may look
for the sum of all multiparton interactions to the hadronic
inelastic cross section. The subset where all multiple parton collisions are
disconnected can be easily summed up in the uncorrelated case\cite{Amet}. 
The result is
the semi-hard hadronic cross section $\sigma_H$, which represents the 
contribution to the hadronic inelastic cross section from events with
at least one semi-hard partonic interaction. The
actual expression is

\begin{eqnarray}
\sigma_H&=&\int d^2\beta\Bigl[1-e^{-\sigma_SF(\beta)}\Bigr]\nonumber\\
 &=&\sum_{n=1}^{\infty}\int d^2\beta{\bigl(\sigma_SF(\beta)\bigr)^n\over n!}
 e^{-\sigma_SF(\beta)}
\end{eqnarray}

\par\noindent
The integration on the impact parameter of the 
hadronic collision, $\beta$, gives the dimensionality to the cross section. The
argument of the integral has the meaning of a Poissonian distribution
of multiple semi-hard partonic interactions with average number depending
on the impact parameter.
\par
The actual value of $\sigma_{eff}$ can be obtained by taking twice the opposite 
of the
second term of the expansion of $\sigma_H$ in powers of multiple collisions, so
the actual value of $\sigma_H$ is related to the value of 
$\sigma_{eff}$ through Eq.(6). The single and the double parton scattering
cross sections are however related to the average number of parton scatterings
and to the second moment. Indeed if one writes the average number of parton
scatterings one obtains:

\begin{eqnarray}
\langle n\rangle\sigma_H&=&\int d^2\beta\sum_{n=1}^{\infty}
{\bigl(n\sigma_SF(\beta)\bigr)^n\over n!}
 e^{-\sigma_SF(\beta)}\nonumber\\
&=&\int d^2\beta \sigma_SF(\beta)=\sigma_S
\end{eqnarray}

\noindent
and for the second moment:

\begin{eqnarray}
\langle n(n-1)\rangle&\sigma_H&=\nonumber\\
=\int d^2\beta& \sum_{n=1}^{\infty}&
{\bigl(n(n-1)\sigma_SF(\beta)\bigr)^n\over n!}
 e^{-\sigma_SF(\beta)}\nonumber\\
=\int d^2\beta& \sigma_S^2&\bigl[F(\beta)\bigr]^2=\sigma_D
\end{eqnarray}

\noindent
The relation between $\sigma_S$ and $\langle n\rangle$ and the
relation between $\sigma_D$ and $\langle n(n-1)\rangle$ just obtained 
do not hold only in the simplest case of the Poissonian distribution
of multiple parton collisions. They
are indeed much more general validity. One can in fact obtain the same
relations in the most general case of multiparton distributions and
keeping moreover into account all semi-hard parton rescatterings\cite{Funct}.
One may therefore write

\begin{equation}
\langle n\rangle\sigma_H=\sigma_S\quad {\rm and}\quad \langle
n(n-1)\rangle\sigma_H=\sigma_D
\end{equation}

\noindent
The effective cross section is defined by the relation

\begin{equation}
\sigma_D={\sigma_S^2\over\sigma_{eff}}\nonumber
\end{equation}

\noindent
one may therefore write 

\begin{equation}
\langle n(n-1)\rangle=\langle n\rangle^2{\sigma_H\over\sigma_{eff}}
\end{equation}
 
\noindent
which implies that in case of an overall Poissonian distribution
of multiple parton collisions one would have $\sigma_{eff}=\sigma_H$.
When the number of parton collisions is very large, in
the simplest uncorrelated case, the distribution is Poissonian
at a fixed value of the impact parameter. The expectation is therefore
that the overall distribution in the number of parton collisions
has a larger dispersion as compared with the Poissonian case. In that
regime $\sigma_{eff}$ is therefore smaller with respect to $\sigma_H$.
The comparison between the actual value of $\sigma_{eff}$ and 
of $\sigma_H$ depends on the 
functional form of $F(\beta)$. In the simplest case 
where $F(\beta)={\rm exp}(-\beta^2/R^2)/\pi R^2$ one obtains a closed
analytic expression for $\sigma_H$:

\begin{equation}
\sigma_H=2\pi R^2\bigl[\gamma+{\rm ln}\kappa+E_1(\kappa)\bigr]
\end{equation}

\par\noindent
where $\gamma=0.5772\dots$ is Euler's constant, $\kappa=\sigma_S/(2\pi R^2)$
and $E_1(x)$ is the exponential integral. In this example the relation with
the hadronic radius $r$ is $R=r\sqrt2$.
For small $\kappa$ one obtains
$\sigma_H\to 2\pi R^2\kappa=\sigma_S$, for large $\kappa$, namely $\sigma_S\to
\infty$, one obtains $\sigma_H\to2\pi R^2\bigl(\gamma+{\rm}ln\kappa\big)$.
Here $\sigma_{eff}=2\pi R^2$. The value of $\sigma_H$
is therefore proportional to the measured value of $\sigma_{eff}$, the 
proportionality factor is slightly dependent on energy and on the cutoff. 
Sensible values of 
the hadron-hadron c.m. energy and of the cutoff give 
values for $\sigma_H$ which are some $30-40\%$ larger with
respect to the value of $\sigma_{eff}$. Different analytic forms for
$F(\beta)$ give qualitatively similar results.
\par\noindent
The effective cross section quoted by CDF is indeed different with
respect to the effective cross section discussed here and in most of the papers
on double parton scatterings. $\sigma_{eff}$ has a simple link with
the overlap of matter distribution in the hadronic collision when 
$\sigma_D$ is obtained from the second moment of the distribution in 
the number of partonic collisions, as discussed above. In the sample
of events with double parton collisions CDF on the contrary has removed
all events where triple parton collisions are present. The correction 
is not a minor one since the fraction of events with triple collisions
is 17\%. In the simplest uncorrelated case just discussed the double parton
scattering cross section measured by CDF would correspond to the
expression

\begin{equation}
\bigl[\sigma_D\bigr]_{CDF}=
\int d^2\beta{\bigl(\sigma_SF(\beta)\bigr)^2\over 2}
 e^{-\sigma_SF(\beta)}
\end{equation}

\noindent
The relation above shows which is the complication introduced by
the requirement of an exclusive cross section. 
In order to make the exclusive selection
of the events with double parton
collisions only, one has to introduce the exponential factor which 
represents the probability of not having any further parton interaction.
This factor, in principle, has a rather complicated dependence on the
overlap of the matter distribution of the two hadrons since, by unitarity,
it is related to the whole series of multiple parton collisions.
The effective cross section quoted by CDF, $(\sigma_{eff})_{CDF}=
14.5\pm1.7^{+1.7}_{-2.3}mb$, refers to the exclusive measurement and
therefore it has to be regarded as an upper bound on the value of the effective
cross section related to an inclusive measurement,
as it has been presently discussed.

The experimental indication is therefore that the effective
cross section is rather small as compared with the naive expectation.
The simplest assumptions
underlying the derivation of Eq.(6) have therefore to be revised.
\par	The main hypothesis which has been done to obtain
the expression for $\sigma_H$ in Eq.(6) is the Poissonian multiparton
distribution. On the other hand
one has to expect correlations between partons as a consequence
of the binding force. While most probably correlations will affect
the $x$ dependence of the multiparton distribution only for finite
values of $x$, and therefore at large rapidities, correlations
in the transverse parton coordinates are present in every kinematical
regime. Indeed the main reason of interest in multiple parton collisions, 
besides the
identification of the process itself, is precisely the measure of
the many-body parton correlations, which is an information on the hadron 
structure
independent on the one-body parton distributions
usually considered in hard processes. 
\par	In the next paragraph we discuss
the most general expression
for the semihard cross section $\sigma_H$, which one obtains by 
\par\noindent
1) assuming
that only two-body parton correlations are present in the many-body parton
distributions and by 
\par\noindent
2) summing all disconnected multiple parton
interactions. 

\section{SEMI-HARD CROSS SECTION AND CORRELATIONS}

\par	At a given resolution, provided by the cut off $p_t^{min}$
that defines the lower threshold for the production of minijets,
one can find the hadron in various partonic configurations. The probability of 
an exclusive $n$-parton distribution,
namely the probability to find the hadron in a configuration
with $n$-partons, is denoted by $W_n(u_1\dots u_n)$.
$u_i\equiv({\bf b}_i,x_i)$ represents the transverse partonic coordinate ${\bf 
b}_i$
and longitudinal fractional momentum $x_i$ while color and flavor variables
are not considered explicitly.
The distributions are symmetric in the variables $u_i$. One defines the 
generating 
functional of the multiparton distributions as:

\begin{eqnarray}
{\cal Z}[J]=&\sum_n&{1\over n!}\int J(u_1)\dots J(u_n)\nonumber\\
&W_n&(u_1\dots u_n)
du_1\dots du_n,
\end{eqnarray}

\noindent
where the dependence on the infrared cutoff $p_t^{min}$ is implicitly 
understood,
and one may introduce also the logarithm of the generating 
functional: ${\cal F}[J]={\rm ln}\bigl({\cal Z}[J]\bigr)$.
The conservation of the probability yields the overall normalization condition

\begin{equation}
{\cal Z}[1]=1.
\end{equation}

\noindent
 One may use the generating functional to derive the many body densities, 
i.e. the inclusive distributions
$D_n(u_1\dots u_n)$:

\begin{eqnarray}
D_1(u)&=&{\delta{\cal Z}\over \delta J(u)}\biggm|_{J=1}
                 ,\nonumber\\
     D_2(u_1,u_2)&=&
                 {\delta^2{\cal Z}\over \delta J(u_1)\delta J(u_2)} 
                  \biggm|_{J=1},\nonumber\\
                 &\dots\nonumber\\
\end{eqnarray}
 
\noindent
The 
many body parton correlations are defined 
by expanding ${\cal F}[J]$ in the vicinity of $J=1$: 

\begin{eqnarray}
{\cal F}[J]&=&\int D(u)[J(u)-1]du\nonumber\\
&+&\sum_{n=2}^{\infty}{1\over n!}
\int C_n(u_1\dots u_n)\bigl[J(u_1)-1\bigr]\dots\nonumber\\
  &\dots&\bigl[J(u_n)-1\bigr]
du_1\dots du_n
\end{eqnarray}

\noindent
Here $D=D_1$ and the correlations $C_n$ describe how much the distribution
deviates from a Poisson distribution, which corresponds in fact to 
$C_n\equiv 0, n\ge 2$.
\par	In the case of hadron-nucleus and nucleus-nucleus collisions a 
systematic use of the AGK cutting rules\cite{Agk} allows one to express 
the total inelastic cross section as a probabilistic superposition
of nucleon-nucleon interaction probabilities\cite{Cs}. The same feature holds 
for the self-shadowing cross sections\cite{Co}. When considering hadron-hadron
collisions as interactions between objects composed with partons, one
can make the assumption that similar relations hold with nucleons
in the place of nuclei and partons replacing nucleons. Of course,
contrary to the nucleon number in the nucleus the parton number is not
fixed. In this respect
semihard parton-parton interactions have to be regarded as a
particular case of self-shadowing interactions\cite{Cn}. 
The semi-hard nucleon-nucleon cross section is then expressed  
as the sum of all the probabilities of multiple parton collisions: 

\begin{equation}
\sigma_H=\int d^2\beta\sigma_H(\beta)
\end{equation}

with

\begin{eqnarray}
&\sigma_H&\!(\beta)=\int\sum_n{1\over n!}
  {\delta\over \delta J(u_1)}\dots 
  {\delta\over \delta J(u_n)}{\cal Z}_A[J]\nonumber\\
  &\times&\sum_m{1\over m!}
  {\delta\over \delta J'(u_1'-\beta)}\dots 
  {\delta\over \delta J'(u_m'-\beta)}{\cal Z}_B[J']\nonumber\\
&\times&\Bigl\{1-\prod_{i=1}^n\prod_{j=1}^m\bigl[1-\hat{\sigma}_{i,j}(u,u')\bigr
]
   \Bigr\}\prod dudu'\Bigm|_{J=J'=0}\nonumber\\ 
\end{eqnarray}

\noindent
where $\beta$ is the impact parameter between the two interacting hadrons
$A$ and $B$
and $\hat{\sigma}_{i,j}$ is the elementary probability for parton $i$ 
(of $A$) to have a hard interaction with parton $j$ (of $B$). 
The semi-hard cross section is constructed summing over all possible
partonic configurations of the two interacting hadrons (the sums over
$n$ and $m$) and, for each configuration with $n$ partons from $A$ and
$m$ partons from $B$, summing over all possible multiple partonic
interactions. This last sum is constructed asking for the 
probability of no interaction between the two configurations
( actually $\prod_{i=1}^n\prod_{j=1}^m[1-\hat{\sigma}_{i,j}]$ ). One
minus the probability of no interaction 
is equal to the sum over all 
semi-hard interaction probabilities.
\par	The presence of multiple parton interactions is
induced by the large flux of partons which is effective at large energies.
The most important contribution to the semi-hard 
cross section, as a consequence, is the contribution of
the disconnected partonic collisions,
namely the interactions where each parton undergoes at most
one semi-hard collision. These are, 
in fact, those multiple partonic interactions that, at a given number of
partonic collisions, maximize the parton flux. Indeed the search
and the observation of the first evidence of multiple semi-hard
parton interactions has been focused to the case of double disconnected 
parton interactions\cite{CDF,Dexp}.
We simplify therefore the problem
by expanding the interaction probability
( the factor in curly brackets ) as sums and by removing all 
the addenda containing repeated indices:

\begin{eqnarray}
\Bigl\{1&-&\prod_{i,j}^{n,m}\bigl[1-\hat{\sigma}_{ij}\bigr]\Bigr\}
\Rightarrow\\ 
  \sum_{ij}\hat\sigma_{ij}&-&{1\over 2!}
  \sum_{ij}\sum_{k\not=i,l\not=j}\hat\sigma_{ij}\hat\sigma_{kl}+
  \dots
\end{eqnarray} 

\noindent
as a result the semi-hard cross section is constructed  with 
multiple disconnected 
parton collisions only, where disconnected refers to the perturbative
component of the interaction. 
Because of the symmetry of the derivative operators in Eq.(19) one can 
replace the expression in Eq.(21) with:

\begin{equation}
nm\hat\sigma_{11}-{1\over 2!}n(n-1)m(m-1)\hat\sigma_{11}\hat\sigma_{22}
+\dots
\end{equation}

\noindent
in such a way that the sums over $m$ and $n$ can be performed explicitly.
As a consequence the cross section at fixed impact parameter, 
$\sigma_H(\beta)$, can be expressed by the operatorial form:

\begin{eqnarray}
\sigma_H(\beta)&=&
  \Bigl[ 1-\exp\bigl(-\delta\cdot\hat{\sigma}\cdot\delta'\bigr)\Bigr]\cr\cr
  &\cdot&{\cal Z}_A[J+1]{\cal Z}_B[J'+1]\Bigm|_{J=J'=0}
\end{eqnarray}

\noindent
We have avoided writing explicitly the variables $u$ and $u'$ and
the functional derivative ${\delta /\delta J(u_i)}$
has been simply indicated as $\delta_i$.   
\par	The form of $\sigma_H(\beta)$ given by Eq.(23) is still too 
complicated to be worked out in its general form, since all possible 
multi-parton correlations are present in ${\cal Z}$. 
Therefore we further simplify the
problem by taking into account two-body parton correlations only.
Our explicit expression for ${\cal F}$ is therefore:

\begin{eqnarray}
{\cal F}[J+1]=&\int& D(u)J(u)du\cr\cr+
                    {1\over 2}&\int& C(u,v)J(u)J(v)dudv
\end{eqnarray}

\noindent
where $D(u)$ is the average number of partons and 
$C(u,v)$ is the two-body parton correlation. 
\par
Either by using techniques of functional
integration or by means of a suitable diagrammatic expansion\cite{Cd} 
one is able to obtain in this case a
closed expression for $\sigma_H(\beta)$:

\begin{equation}
\sigma_H(\beta)=1-\exp\Bigl[-
\frac{1}{2} \sum_n a_n -\frac{1}{2} \sum_n b_n /n\Bigr]
\end{equation}
 
\noindent
where  $a_n$ and $b_n$ are functions of the impact parameter $\beta$ and are 
given by

\begin{eqnarray}
a_n&=&\!\int D_A(u_1)\hat \sigma (u_1,u'_1)\times\cr\cr 
 &\times&C_B(u'_1-\beta,u'_2-\beta)\hat 
 \sigma (u'_2,u_2) C_A(u_2,u_3)\dots\cr\cr 
 &\dots& D_B(u'_n-\beta) 
 \prod du_i du'_i 
\end{eqnarray}

\begin{eqnarray}
b_n&=&\!\int C_A(u_n ,u_1)\hat \sigma (u_1,u'_1)\times\cr\cr 
  &\times&C_B(u'_1-\beta,u'_2-\beta)\hat 
 \sigma (u'_2,u_2) \dots\cr\cr
 &\dots&  C_B(u'_{n-1}-\beta,u'_n-\beta)\hat 
 \sigma (u'_n,u_n) \cr\cr
 &\prod& du_i du'_i\,.
\end{eqnarray}

The actual expression for $a_n$ holds 
for $n$ odd. When $n$ is odd one may also have the symmetric case,
where the expression begins with $D_B$ and 
ends with $D_A$. When $n$ is even the initial and final 
distribution are either both $D_A$ or both $D_B$. In the definition of $b_n$
$n$ is always even, so that one of
the ends is $A$ and the other is $B$. One may notice
that, at a given order in the number of partonic interactions, one can obtain a 
term of kind {\it a} from a term
of kind {\it b} by replacing one $C$ with a pair of $D$'s. 
The operation can be done in $n$
ways. The combinatorial meaning of the $1/n$ factor multiplying 
each term of kind {\it b} in Eq.(25) is then 
understood. The factor $1/2$ in Eq.(25) is the consequence of the symmetry
between $A$ and $B$.
\par
 The cross section is given by an integral on the
impact parameter of the interaction probability, $\sigma_H(\beta)$, that is 
expressed as
one minus the probability of no interaction. The probability of no interaction
is given by the negative exponential of the sum over all possible different
connected structures, namely all structures of kind $a_n$ and of kind $b_n$.
With our approximations, Eq.(21) and Eq.(24), these are in fact all possible 
connected
structures which can be built with the average numbers $D_{A,B}$,
the two-body correlations $C_{A,B}$ and the interaction $\hat{\sigma}$ .
Expanding the exponential, the cross section can then be expressed as the 
sum over all possible structures, both connected and disconnected.
\par	One will notice that, when no correlations are present, all terms
of kind {\it b} disappear and only the first of the terms of kind {\it a},
namely $D_A\hat{\sigma}D_B$ is left. 
In that limit the cross section is given simply by:

\begin{equation}
\sigma_H=\int d^2\beta\bigl\{1-e^{-\langle n(\beta)\rangle}\bigr\}
\end{equation}

\noindent
where

\begin{equation}
\langle n(\beta)\rangle=
\int D_A(u-\beta)D_B(u')\hat{\sigma}(u,u')dudu'
\end{equation}

\par\noindent
which corresponds to the Poissonian distribution discussed in the
introduction.

\section{TWO DIFFERENT QUALITATIVE FEATURES OF THE CORRELATION TERM}

The small value of $\sigma_{eff}$, the dimensional parameter
characterizing double parton scatterings, which has been measured
recently by CDF, is an indication that two-body parton correlations,
in the many-body parton distribution of the proton,
are likely to be sizable.
In the case of an uncorrelated many-body parton distribution,
the value of $\sigma_{eff}$ puts a constraint on the
range of possible values of $\sigma_H$, the semi-hard contribution to the
hadronic inelastic cross section. The actual measured value of $\sigma_{eff}$
would give rise to values of $\sigma_H$ of the order of $\sigma_{inel}/2$
also at very large c.m. energies, where one would rather 
expect $\sigma_H\simeq\sigma_{inel}$.
The experimental evidence is also that, in the $x$ region accessible
experimentally namely at small $x$ values, the correlation in fractional
momenta is not a large effect. 
\par
$\sigma_H$ can be worked out rather explicitly when only
two-body parton correlations are included
in the many-body parton distributions and when each parton can have at
most one semi-hard interaction. Two qualitatively different features can be
present
in the two-body parton correlation, and both change the relation
between $\sigma_H$ and $\sigma_{eff}$ with respect to the uncorrelated case: 

1- The distribution in the
number of partons is not any more a Poissonian, although the dependence on the 
kinematical variables of the different partons is factorized.

2- The overall distribution in the number of partons, obtained
after integrating on the partonic kinematical variables, is a Poissonian
but the dependence on the partonic kinematical variables is not factorized,
in this case the two-body parton correlation integrates to zero.
\par\noindent
The general case is obviously a combination of the two possibilities.
We point out however that
both cases separately can give rise to a small value of 
$\sigma_{eff}$ while keeping the 
value of $\sigma_H$ close to $\sigma_{inel}$.
\par
One can work out explicitly the expression for the semi-hard cross section
in Eq.(25) considering explicit examples\cite{Cu}. The general result 
is however that
the critical value of the impact parameter $\beta_c$, which gives the size
to the cross section $\sigma_H$, is the value which makes small the argument
of the exponential in the expression of $\sigma_H(\beta)$. The detailed 
dependence of the argument of the exponential at $\beta<\beta_c$ is not
of great importance for the determination of $\sigma_H$
when, for $\beta<\beta_c$, the argument of the exponential is
already large: $\sigma_H$ is obtained by integrating the
probability of having at least one semi-hard interaction. When the
probability to have at least one semi-hard interaction is close to one, 
the contribution to the integral is very
similar for events with the same impact parameter and with different
but large average number of partonic collisions.
\par
The critical value of the impact parameter which gives the size 
to $\sigma_H$ is therefore determined by the argument of the exponential
at the edge of the interaction region. The dominant
contribution at the edge is due to the single scattering term, since
higer order collision terms are important when the density of 
overlapping matter of the two hadrons is large. This is precisely
the argument of the exponential in the uncorrelated case and the
consequence is that the resulting value of $\sigma_H$ is not 
very different with respecy to the uncorrelated case.
Actually $2\pi R^2$ as discussed in the introduction.
\par
The correlation term is on the contrary able to change sizably
the effective cross section. One may modify the number distribution, 
without introducing non-factorized
two-body correlations in ${\bf b}$, by using by using for example
the factorized expression 

\begin{equation}
C(u,u')=-\lambda D(u) D(u')
\end{equation}

\par\noindent
One obtains in this case the relation\cite{Cu}

\begin{equation}
\sigma_{eff}={2\pi R^2\over (1+\lambda)^2}
\end{equation}

\noindent
If one introduces a correlation term which does not modify
parton number distribution and which therefore integrates to zero,
the double scattering cross section is incresed, with respect to
the uncorrelated case, by an additive term corresponding to
the convolution of two correlations\cite{Cu}:  

\begin{eqnarray}
{1\over\sigma_{eff}}&=&{1\over2\pi R^2}+\cr\cr
&+&\int C({\bf b}, {\bf b}')
C({\bf b}-\beta, {\bf b}'-\beta)\cr\cr
&\times&d^2bd^2b'd^2\beta
\end{eqnarray} 

\par
A qualitative feature is that in both cases one obtains a 
value of $\sigma_{eff}$ which may be sizably smaller
with respect to $2\pi R^2\simeq\sigma_H$.
While, on the other hand, nothing prevents the value of $\sigma_H$ from
being close to the value of $\sigma_{inel}$. 
The smaller value of $\sigma_{eff}$, with respect to the 
expectation of the uncorrelated case, is
rather generally
associated with the increased dispersion of the distribution
in the number of partonic collisions: In the case of no correlations
the distribution is strictly Poissonian when the impact parameter is fixed.
When correlations are introduced the
distribution in the number of parton collisions, at fixed $\beta$,
is not Poissonian any more
and the natural consequence is that the dispersion in the number of collisions 
is increased.
\par
The indication from the measure of the rate of double parton scatterings
is therefore that two-body parton correlations are likely to be important
while, unfortunately, one cannot say much about dynamical quantities, like the 
the correlation
length. Useful observables to be measured, 
in order to get some more insight into the problem,
would be the semi-hard cross section $\sigma_H$ and the triple parton
scattering cross section. 
The measure of $\sigma_H$, in association with $\sigma_{eff}$, would 
help considerably in clarifying the size of the effect induced by the presence
of the two-body parton correlations: All present considerations 
are based on the prejudice that $\sigma_H$ should have a value
rather close to the value of $\sigma_{inel}$. 
\par\noindent
The measure of triple and of higher order parton scatterings would 
give important constraints on models of the many body parton
distributions. For example if only lower order correlations where important
one would be able to fix all the correlation parameters.
\par
While a lot of effort has been put in the study of the proton structure
as a function of the momentum fraction $x$, one should keep in mind that
the distribution of
partons depends on three degrees of freedom, the momentum
fraction $x$  and the transverse parton coordinate ${\bf b}$.
The measure of multiple parton collisions is the essential tool
which allows us 
to learn on the parton structure of the proton in transverse
plane.


\begin{thebibliography}{9}

\bibitem{Double} C. Goebel, F. Halzen and D.M. Scott, {\it Phys. Rev.}
{\bf D22}, 2789 (1980);
N. Paver and D. Treleani, {\it Nuovo Cimento} {\bf A70},
215 (1982); B. Humpert, {\it Phys. Lett.} 
{\bf B131}, 461 (1983); B. Humpert and R. Odorico, {\it ibid} {\bf 154B}, 211
(1985); T. Sjostrand and
M. Van Zijl, {\it Phys. Rev.} {\bf D36}, 2019 (1987).
\bibitem{CDF} F. Abe et al. (CDF Collaboration), submitted to 
{\it Phys. Rev.} {\bf D} April 14, 1997;
FERMILAB-PUB-97/094-E.
\bibitem{Dexp} T. Akesson et al. (AFS Collaboration), {\it Z. Phys.} {\bf C34}
163 (1987); J. Alitti et al. (UA2 Collaboration), {\it Phys. Lett} {\bf B268}
145 (1991); F. Abe et al. (CDF Collaboration), {\it Phys. Rev.} {\bf D47}
4857 (1993).
\bibitem{Amet} Ll. Ametller and D. Treleani, {\it Int. J. Mod. Phys.} {\bf A3},
521 (1988). 
\bibitem{Funct} G. Calucci and D. Treleani, {\it Nucl. Phys.} 
{\bf B} (Proc. Suppl.) 18C, 187 (1990) and 
{\it Int. J. Mod. Phys.} {\bf A6}, 4375 (1991).
\bibitem{Agk} V. Abramovskii, V.N. Gribov and O.V. Kancheli, {\it Yad. Fiz.}
{\bf 18}, 595 (1973) [{\it Sov. J. Nucl. Phys.} {\bf 18}, 308 (1974) ];
I.G. Halliday and C.T. Sachrajda, {\it Phys. Rev.} {\bf D8}, 3598 (1973);
J. Koplik and A.H. Mueller, {\it Phys. Lett.} {\bf 58B}, 166 (1975);
L.D. Mc Lerran and J.H. Weiss, {\it Nucl. Phys.} {\bf B100}, 329 (1975);
L. Bertocchi and D. Treleani, {\it J. Phys.} {\bf G3} 147 (1977);
V.M. Braun and Yu.M. Shabelski, {\it Int. J. Mod. Phys.}
{\bf A3}, 2417 (1988);
G. Calucci and D. Treleani {\it Phys. Rev.} {\bf D49}, 138 (1994);
{\bf D50}, 4703 (1994);
J. Bartels and M. W\"{u}sthoff, {\it Z. Phys.} {\bf C66}, 157 
(1995).
\bibitem{Cs} A. Capella and A. Krzywicki, {\it Phys. Lett.} {\bf 67B}, 84 
(1977);
{\it Phys. Rev.} {\bf D18}, 3357 (1978).
\bibitem{Co} R. Blankenbecler, A. Capella, C. Pajares, J. Tran Thanh Van and
A.V. Ramallo, {\it Phys. Lett.} {\bf 107B}, 106 (1981).
\bibitem{Cn} D. Treleani, {\it Int. J. Mod. Phys.} {\bf A11} 613 (1996).
\bibitem{Cd} G. Calucci and D. Treleani, {\it Nucl. Phys.} 
{\bf B} (Proc. Suppl.) 18C, 187 (1990) and 
{\it Int. J. Mod. Phys.} {\bf A6}, 4375 (1991).
\bibitem{Cu} G. Calucci and D. Treleani, hep-ph/9708233 to be published on
{\it Phys. Rev.} {\bf D}.

\end{thebibliography}
\end{document}